\documentclass{mn2e}
\usepackage{natbib,amssymb,amsmath}
\usepackage{graphicx}
\newcommand{\degree}{\ensuremath{^\circ}}
\numberwithin{equation}{section}
\title[Analysing the Transverse Structure of the Relativistic Jets of AGN]
{Analysing the Transverse Structure of the Relativistic Jets of AGN}

\author[Murphy, Cawthorne \& Gabuzda]{E. Murphy$^{1}$, T. V. Cawthorne$^{2}$ \& D. C. Gabuzda$^{1}$\\
$^{1}$Physics Department, University College Cork, Cork, Ireland\\
$^{2}$School of Computing, Engineering and Physical Science, University of 
Central Lancashire, Preston, England} 

\begin{document}
\date{}
\maketitle

\begin{abstract}
This paper describes a method of fitting total intensity and polarization 
profiles in VLBI images of astrophysical jets to profiles predicted by 
a theoretical model. As an example, 
the method is used to fit profiles of the jet in the Active Galactic Nucleus 
Mrk~501 with profiles predicted by a model in which a cylindrical jet of
synchrotron plasma is threaded by a magnetic field with helical and disordered 
components. This fitting yields model Stokes $Q$ profiles that agree with 
the observed profiles to within the $1-2\sigma$ uncertainties; 
the $I$ model and observed profiles are overall not in such good 
agreement, with
the model $I$ profiles being generally more symmetrical than the observed
profiles. Consistent fitting results are obtained for
profiles derived from 6~cm VLBI images at two distances from the core,
and also for profiles obtained for different wavelengths at a single location
in the VLBI jet.  The most striking success of the model is its ability
to reproduce the spine--sheath polarization structure observed across the
jet.
Using the derived viewing angle in the jet rest frame, $\delta^{\prime} \simeq 
83^{\circ}$, together with a superluminal speed reported in the literature, 
$\beta_{app} = 3.3$, yields a solution for the viewing angle and velocity of 
the jet in the observer's frame $\delta \simeq 15^{\circ}$ and $\beta\simeq 
0.96$.  Although these results for Mrk501 must be considered tentative, 
the combined analysis of polarization profiles and apparent component speeds 
holds  promise as a means of further elucidating the magnetic field structures 
and other parameters of parsec-scale AGN jets.
\end{abstract}

\section{Introduction}

At radio wavelengths, the jets of active galactic nuclei (AGNs) emit synchrotron 
radiation which is characterised by appreciable linear polarization,  the plane 
of polarization being perpendicular to the sky projection of the source magnetic 
field. The polarization structure of  jets provides information about the 
structure of their magnetic fields, which in turn influence their evolution and 
emission properties. Magnetic field structures in AGN jets are of great 
importance,  for instance, they affect jet stability. A knowledge of magnetic 
field structure is required in order to translate radio images into jet 
structure and  also provides constraints on jet formation models.  Despite much 
observational effort  the magnetic field structures of AGN are not yet well 
understood. 

Three types of observational results inform our present thinking about the
magnetic field structures in parsec-scale jets. First, there is a tendency for
polarization angles to lie either parallel or perpendicular
to the jet. A tendency for BL Lac objects and quasars to display magnetic
polarization directions respectively perpendicular and parallel to the
jets has long been known (e.g. Gabuzda et al. 1992; Cawthorne et al. 1993),
and modern surveys (Lister \& Homan 2005) have confirmed
these general trends. In quasars, there is a broad distribution of misalignments
but such differences are often reduced when Faraday rotation is taken into 
account (e.g. Hutchison et al. 2001). This type of polarization structure 
requires an axisymmetric magnetic field; a helical magnetic field is one 
example, although the production of apparent jet magnetic fields that are 
either parallel or perpendicular to the jet direction is not a unique property 
of helical fields.

Secondly, transverse Faraday rotation gradients have been reported 
across a number of AGN jets (e.g. Asada et al. 2002, Gabuzda et al.
2004). These results, though somewhat controversial (e.g. Zavala \& Taylor 
2010), suggest the existence of a toroidal magnetic field component.

Thirdly, a significant number of AGN jets possess obvious asymmetries in
total intensity and linear polarization or other transverse structures 
that are reminiscent 
of those revealed in the helical field simulations of Laing (1981). 
Note that the presence of transverse Faraday rotation gradients 
does not provide any information about whether 
or not the poloidal field component is ordered. Generally, speaking, it is 
only the asymmetry of the transverse intensity and polarization profiles 
that can distinguish observationally between a helical field (with an ordered 
poloidal field component) and a toroidal field (with a disordered poloidal 
field component).

These observational results lead to the main question addressed in this 
paper: can a helical magnetic field
explain the observed intensity and
polarization profiles of parsec-scale AGN jets? 

Of course, asymmetries in the transverse profiles 
could also be attributed to physical asymmetries, 
such as pressure gradients or other forms of jet asymmetry. 
However, a helical magnetic field 
threading the jet of an AGN could potentially 
describe the observations without requiring special conditions in the jet 
environment. The main advantage of such models is that they can produce  
the observed transverse structures while avoiding physical 
asymmetries which might cause the jet to deflect or possibly destabilise.

In recent years, Laing et al. (e.g. 2006a, 2006b, 2008) have also investigated 
whether ordered magnetic fields are present in the jets of FR1 radio galaxies on 
scales much larger than those investigated in this paper.  The synchrotron 
emission from various magnetic-field configurations was calculated based on 
three-dimensional models, with the aberration calculations done via numerical 
integration for the correct spectral indices. These theoretical results were 
compared to the observed total intensity and polarized emission from several 
FR1 sources, via fitting to two-dimensional brightness distributions containing 
more than 1000 independent pixels. This work indicates that the 
polarization data for larger-scale jets are compatible with, but do not require, 
the presence of an ordered toroidal component; at least some kiloparsec-scale 
observations seem to be inconsistent with the presence of a strong, ordered,
large-scale poloidal 
component, as is expected on theoretical grounds (e.g. Begelman et al. 1984),
suggesting that large-scale helical fields may be ruled out in some 
cases. However, a transition from a helical field configuration on parsec scales
to an ordered toroidal (+ disordered poloidal) field configuration on kiloparsec
scales is plausible, so that a lack of support for the presence of a helical field
component on kiloparsec scales need not imply the same for parsec scales. 

The present paper describes an approach to fitting model jet 
total intensity and polarization profiles to observed profiles taken across
VLBI jets. The observed jet profiles for the BL Lac object Mrk501 are 
considered as an example, using a model with a cylindrical jet threaded by a 
helical magnetic field, as in Laing (1981).  
The method presented here provides a new means to search for evidence for
a helical magnetic-field component in parsec-scale jets, that does not depend
on measurements of Faraday rotation gradients, whose reliability can be 
difficult to determine in some cases. The method
yields estimates of the viewing angle and helical-field pitch angle in the
jet rest frame, which can be used to derive the corresponding quantities in
the observer's frame when combined with observations of the apparent speeds
of jet components.

\section{Helical Magnetic Field Models}

Laing (1981)  investigated three different helical magnetic field models, 
which can be used to predict  the total intensity and 
polarization profiles across a jet using only two parameters, the helical 
pitch angle $\gamma$ and the jet's angle to the line of sight  $\delta$, both 
defined in the co-moving frame; it is assumed that the velocity of the jet
remains constant across the jet width. In applying these models to the 
parsec-scale jets of several AGNs, Papageorgiou (2005) found the 
best agreement between observed and model profiles arose for the third 
of these models, in which a helical magnetic field of constant pitch angle 
threads a cylindrical jet.  It is therefore this model that is the focus of
the present work.  

Because the quantity $\sin{2\chi}$ is antisymmetric, where $\chi$ is the
electric vector position angle (EVPA), the
contributions to the integral of $U$ along the line of sight made by the far
and near sides of the helical field cancel, so that this integral is zero;
therefore, the polarization distribution across the cylinder corresponds
fully to $Q$ (Laing 1981), with

\[\chi = \left\{ 
\begin{array}{l l}
  90\degree & \quad \mbox{if $Q > 0$}\\
 0\degree & \quad \mbox{if $Q < 0$}\\ \end{array} \right. \]

In other words, the integrated magnetic vector position angle is transverse 
for $Q > 0$ and longitudinal for $Q < 0$.
The derivation of Stokes $I$ and $Q$ for this 
model can be found in Laing (1981). Note that there is a typographical error
in his final equation for $Q(x)$ in his Appendix A --- the sign of the second 
term is incorrect, and the correct expression is
\begin{multline}
Q(x) = \frac{p_{0}f}{\sin{\delta}}[a(\sin^2{\gamma}-\cos^2{\gamma}\sin^2{\delta})-
b\sin{2\gamma}\sin{2\delta}\\
- c\sin^2{\gamma}(1+\cos^2{\delta})].
\end{multline}
 
The analysis of Papageorgiou (2005) showed that this helical-field 
model produced  profiles that are considerably more strongly polarized than 
observed.  The model was therefore modified by introducing a disordered 
(tangled) magnetic field component (see, e.g.,  Burn 1966). 
This requires a third model parameter, the 
degree of entanglement, $f$,  defined as the fraction of the magnetic field 
energy density in tangled form:
\begin{equation}
 \frac{\langle B_T^2\rangle}{\langle B_H^2\rangle} = \frac{f}{1-f}
\end{equation}
\\
where $\langle B_T^2\rangle$ and $\langle B_H^2\rangle$ are proportional to 
the magnetic energy densities of the tangled and helical magnetic field 
components.
Increasing the degree of entanglement in the field 
(i.e. increasing $f$) reduces the  degree of asymmetry  of the total intensity 
profiles predicted by the model in addition to decreasing the degree of 
polarization.

Making the same assumptions as Laing (1981) (spectral index $\alpha$ = 1, 
where $S\propto \nu^{-\alpha}$, and constant electron density throughout 
the cylinder), profiles for total intensity ($I(x)$) and polarized intensity 
($Q(x)$) across the jet can be derived analytically:
\begin{equation}
I(x) =\frac{2}{3}C B_{T}^2 \frac{\sqrt{R^2-x^2}}{\sin \delta} + (1-f)I_L(x)
\end{equation}
\begin{equation}
Q(x) = (1-f)Q_L(x)
\end{equation}
where $I_L(x)$ and $Q_L(x)$ are the line-of-sight integrated Stokes parameters 
for Laing's helical-field model, $x$ is distance from the jet axis (projected 
on the sky) 
and $R$ is the radius of the jet cylinder. For $Q$ positive and negative, the 
EVPA is parallel to and normal to the jet direction, respectively.  
\begin{figure*}
\includegraphics[width=1\textwidth]{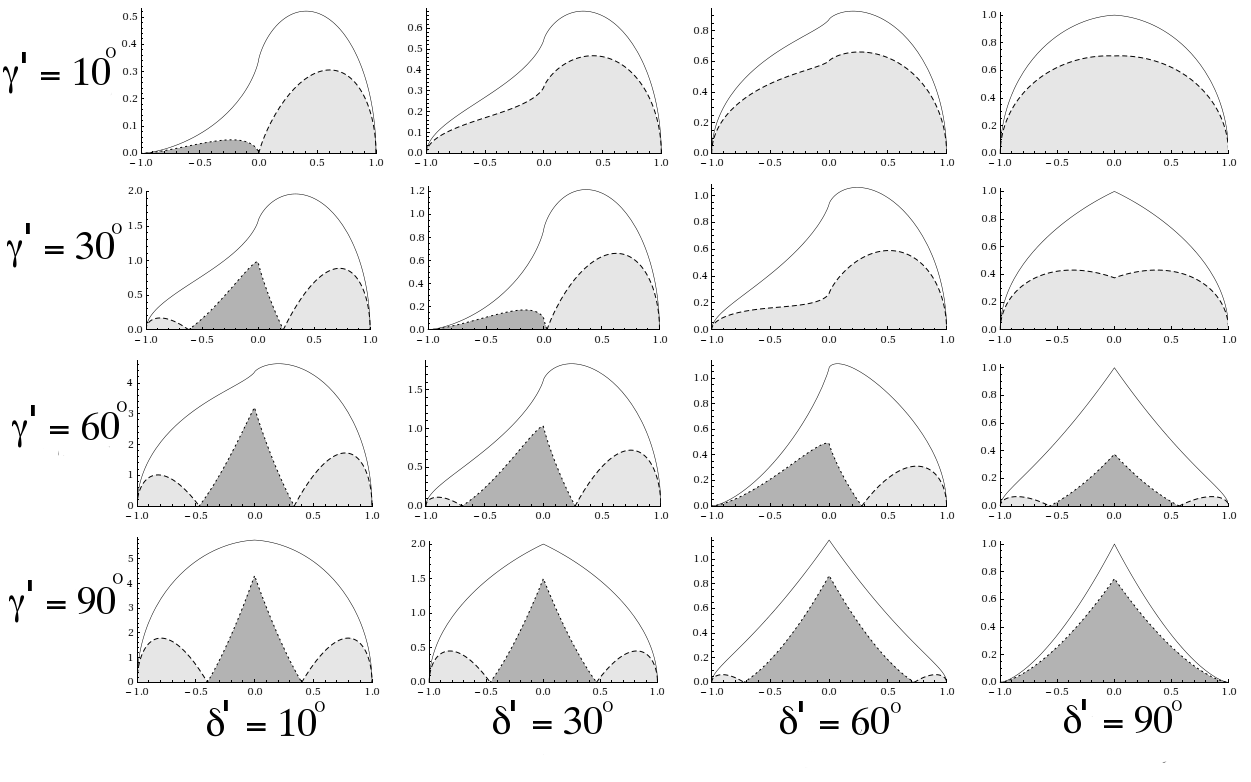}
\caption{Transverse structure produced by the model considered, for various 
viewing angles, $\delta^{\prime}$, and helix pitch angles, $\gamma^{\prime}$. 
Solid lines 
correspond to total intensity, dark grey regions to longitudinal polarization 
(EVPAs aligned with the jet) and light grey regions to transverse 
polarization (EVPAs orthogonal to the jet).}
\label{ProfileTable}
\end{figure*}

The model described so far describes only the emission mechanism in the
rest frame of the jet. If the jet is relativistic then, although the
overall levels of flux density will be changed 
by the Doppler effect and
aberration, the transverse profiles will remain unchanged provided
that the jet's velocity is constant across its width.

The Stokes parameters $I$ in the observer's frame (unprimed) are 
related to those in the rest frame of the jet (primed) by
\begin{equation}
I(\delta) = I^{\prime}(\delta^{\prime})D^a
\end{equation}
where $\delta$ is the viewing angle in the observer's frame, $\delta^{\prime}$ 
is the viewing angle in the rest frame of the jet, $D = 
[\Gamma (1-\beta\cos\delta)]^{-1}$ is the Doppler factor ($\Gamma$ is the
Lorentz factor and $\beta$ the jet
velocity divided by the speed of light) and the value of $a$ is $2+\alpha$, 
where $\alpha$ is the spectral index (Scheuer and Readhead, 1979; Rybicki
and Lightman 1979, Chapter 4). Strictly speaking, Equation (2.5) refers to 
the intensity, $I_{\nu}$, from a continous, optically thin jet, but also
applies to the flux density per unit length from a cylindrical jet.
The Stokes parameters $Q$ and $U$ transform
in precisely the same way.

If the velocity is constant across the width of the cylindrical jet, the 
Doppler factor is the same for every point in the jet. Therefore, in this 
case, the Doppler factor will affect only the overall level of the intensity, 
not the shapes of the $I$, $Q$ or $U$ profiles. More precisely, the jet 
profiles $I(x)$, $Q(x)$ and 
$U(x)$ viewed at angle $\delta$ in the observer's frame will be identical in
shape to the profiles $I^{\prime}(x)$, $Q^{\prime}(x)$ and $U^{\prime}(x)$ 
viewed at angle 
$\delta^{\prime}$ in the jet rest frame.  Because the fits were obtained by
comparing only the shapes of the $Q$ and $I$ profiles, without using the 
absolute intensities of these 
profiles, the resulting fitted profiles correspond directly to the 
pitch angle and the viewing angle in the rest frame of the jet. Further, 
the profiles of the polarization angle and the fractional polarization 
will be identical in the two frames. To be explicit, a prime is used to 
denote quantities in the rest frame of the jet, and absence of a prime 
to denote quantities in the rest frame of the observer.

Although the viewing angles in the observer's frame are likely to be 
small, of the order of $1/ \Gamma$, where $\Gamma$ is the jet Lorentz factor,
the rest-frame viewing angles will be much larger, and it is these rest-frame 
values that we have derived from our profile fits and quote in the 
text and tables below. 

This is essentially the approach taken by Claussen-Brown, Lyutikov and Kharb (2009), 
except that they use a force-free magnetic field 
configuration (and a Gaussian jet profile) defined in the jet's rest frame,
whereas a simple helical field threading a 
uniform jet has been used here. This choice was made because 
the objective of this work is precisely to look for observational signatures
of a helical field component, rather than to look for additional features 
that will inevitably be present in more sophisticated models. (Note that 
Claussen-Brown et al. (2009) quote observer-frame viewing angles, but these can 
only be given in terms of the (unknown) Lorentz factor.)

Zakamska, Begelman and Blandford (2008) consider a relativistic MHD jet with a 
toroidal magnetic field. The main difference between this model and Laing's
is the existence of a poloidal field component in Laing's. A purely
toroidal field configuration can give rise to `spine--sheath' polarization 
structure and systematic transverse Faraday-rotation gradients, but not to 
asymmetric transverse intensity and polarization structure; therefore, it is 
of interest to try to investigate such asymmetric structures using the model 
presented here. 

Broderick and Loeb (2009) and Broderick and McKinney (2010) consider 
theoretical simulations of transverse Faraday rotation measure gradients 
produced by helical jet magnetic fields. The latter simulations, in particular, 
directly demonstrate the generation of a helical field and associated 
Faraday-rotation gradients due to the combination of the rotation of the 
jet base and the jet outflow. These fully relativistic simulations 
show that the resulting transverse Faraday-rotation gradients 
can sometimes prove to be more complex than is predicted by simple 
helical-field models; however, their origin remains the toroidal component 
of an essentially helical jet magnetic field.
 
While detailed models for relativistic jets and their magnetic fields have 
been investigated in these and other studies, the least model-dependent way
to seek evidence for a helical field component is to
is to use results for a purely helical field; this amounts to focusing 
on the basic features of transverse jet profiles rather than aiming to 
describe them in detail in the face of many unknown parameters.  If 
high-quality, high-resolution data become available for jet profiles, then 
it may become feasible to investigate more detailed models.

\section{Model Predictions}

For convenience, the most notable features predicted by this model (Laing 1981), 
visible in the sample grid of profiles of total and polarized intensity for a range of 
$\gamma^{\prime}$ and $\delta^{\prime}$ values given in Fig.~\ref{ProfileTable},
are summarised here.
\begin{enumerate}
\renewcommand{\theenumi}{(\arabic{enumi})}
 \item{Except for purely toroidal magnetic fields ($\gamma^{\prime} = 90\degree$) 
or viewing angles (in the rest frame of the jet) perpendicular to the cylinder 
axis ($\delta^{\prime} = 90\degree$), the distributions of total and polarized 
intensity are usually asymmetric. These asymmetries arise because the 
change in the sky-projected magnetic-field direction occurs most rapidly along 
the helical field line on the side of the jet where the angle between the field 
and the line of sight is smallest. This results in a greater level of 
polarization cancellation along the line of sight on one side of the jet than 
the other. The profile asymmetries are clearly seen in cases where neither 
the helical pitch angle nor the viewing angle equals 90$\degree$ in 
Fig.~\ref{ProfileTable}.}
 \item{Displacements between the total intensity profile maxima and polarized 
intensity profile maxima are common.}
 \item{The fractional polarization varies considerably across the profile.}
 \item{The polarized intensity distribution can have one, two or three local 
maxima, and the orientation of the projected magnetic field can be either 
longitudinal, transverse or a combination of both within a given profile.}
\end{enumerate}
Further examination of the polarization profiles produced by this model shows 
that 4 different magnetic field distributions are possible, as 
can be seen in Fig.~\ref{ProfileTable}.
\begin{enumerate}
\renewcommand{\theenumi}{(\arabic{enumi})}
 \item{Longitudinal all across the jet; e.g., $\gamma^{\prime} = 10\degree$ and 
$\delta^{\prime} = 30\degree$}
 \item{Longitudinal on one side and transverse on the other; e.g., 
$\gamma^{\prime} = 30\degree$ and $\delta^{\prime} = 30\degree$}
 \item{Longitudinal at the edges and transverse at the centre; e.g., 
$\gamma^{\prime} = 30\degree$ and $\delta^{\prime} = 10\degree$}
 \item{Transverse all across; e.g., $\gamma^{\prime} = 90\degree$ and 
$\delta^{\prime} = 90\degree$}
\end{enumerate}
Configuration 2 only occurs when $\gamma^{\prime} = \delta^{\prime}$ and 
configuration 4 only occurs when $\gamma^{\prime}=\delta^{\prime}=90\degree$. 
These special cases would not often be expected in nature. However,  the effects 
of finite resolution, in which the true jet profile is convolved with a 
Gaussian beam, increase the range of parameter values for which configurations 
2 and 4 are observed [See Figure 2, adapted from Papageorgiou (2005)].

\begin{figure}
\centering
\includegraphics[width=1\columnwidth]{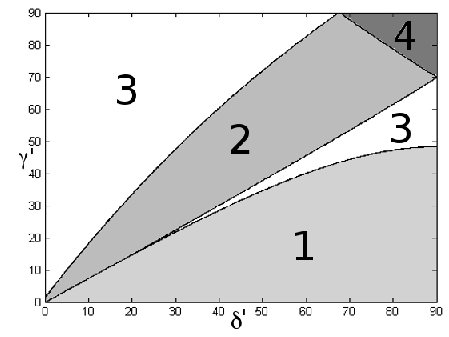}
\caption{Effect of finite resolution on the observed magnetic field 
configurations. Convolution was performed using a Gaussian beam with a FWHM 
one quarter the size of the intrinsic jet width. Region numbers correspond 
to the configuration types listed in Section 3.}
\label{fig:2}
\end{figure}

\begin{figure}
\centering
\includegraphics[width=1\columnwidth]{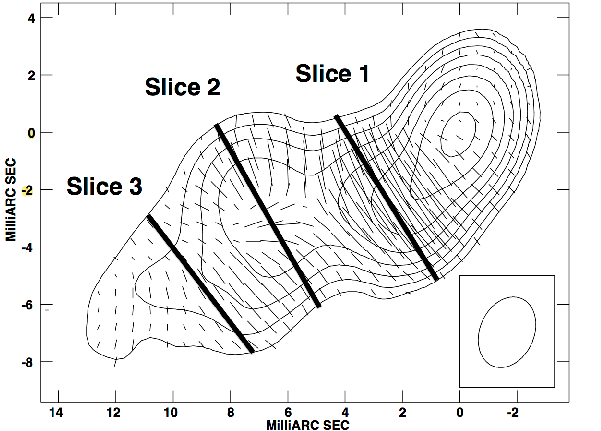}
\caption{6cm Polarization Map of Mrk501 for February 1997 (Pushkarev et al. 
2005). The polarization sticks are proportional to the polarized intensity.  
The three lines across the jet show the transverse slices that were analysed. 
The peak intensity is 0.52 Jy/beam, and the contour levels
are 0.60, 1.25, 2.5, 5.0, 10.0, 20.0, 40.0, and 80.0\% of the peak.} 
\label{fig:3}
\end{figure} 

\section{Comparison Method}

\begin{table*}
\centering
\caption{Best fit Parameters for 6cm Mrk501 Slices (Epoch February 1997)}
\begin{tabular}{@{}lccccc}
\hline
Slices & $\gamma^{\prime}$ & $\delta^{\prime}$ & $f$ & Intrinsic Jet Width\\
\hline
Slice 1 & $41^{\circ}\pm 3^{\circ}$& $81^{\circ}\pm 3^{\circ}$ & $0.70\pm 0.05$ & 3.1 mas\\
Slice 2 & $53^{\circ}\pm 1^{\circ}$ & $80^{\circ}\pm 2^{\circ}$ & $0.40\pm 0.05$ & 5.7 mas\\
Slice 3 & $47^{\circ}\pm 1^{\circ}$ & $90^{\circ}\pm 2^{\circ}$ & $0.00\pm 0.10$ & 4.8 mas \\
\hline
\end{tabular}
\end{table*}

Papageorgiou (2005) did not carry out any formal model fitting, and 
matched the observed and model profiles using a number of simple criteria. 
This approach was time consuming, and did not necessarily result in the best
fit in a `least-squares' sense. This technique has been improved upon here
by generating a database of theoretical profiles to enable a quantitative 
comparison of the observed and model profiles. To generate such a database 
an estimate of  the intrinsic jet width must first be made. This  was done 
by generating a series of jet  profiles of increasing width, which were then 
convolved with the observing beam. The intrinsic jet width was then taken to 
be that for which the convolved $I$ profile best matched that observed. Both 
Gaussian and top-hat model profiles produced similar results. Since the jet 
width determined in this way is only an estimate, several trial values were 
used, as is described in Section 5.

The model transverse profiles were generated for a jet with angular width 
determined as described above and convolved with a Gaussian beam corresponding 
to the observing beam, varying the values of $\delta^{\prime}$, 
$\gamma^{\prime}$ and $f$. The best-fit model was taken to be the model giving 
the smallest residual $\chi^2$ between the observed and theoretical profiles:
\begin{equation}
\chi^2 =\frac{Q_{max}}{I_{max}}\displaystyle\sum\limits_{n=1}^N 
\frac{(I_n-I_n^0)^2}{\sigma_I^2}+\displaystyle\sum\limits_{n=1}^N 
\frac{(Q_n-Q_n^0)^2}{\sigma_Q^2}
\end{equation}
where $N$ is the total number of data points, $I_n$  and $Q_n$ are the  
$n$th observed total and polarized intensity datapoints respectively, 
$I_n^0$ and $Q_n^0$ are the the $n$th model total and polarized intensity 
model values respectively, and $I_{max}$ and $Q_{max}$ are the maximum values
of the observed total and polarized intensity respectively. The $I$ profiles
are downweighted compared to the $Q$ profiles by the factor $Q_{max}/I_{max}$. 
This factor is close to 0.10 for most of the profiles, and effectively gives
the $Q$ and $I$ profiles comparable weights in the fitting, while also ensuring
that the polarization structure of the best fit matches the observed polarization
structure. In effect, we are 
ignoring the higher signal-to-noise ratios of the $I$ data, which is justifiable 
because the real errors that dominate the $I$ image are not due to noise, but 
instead to mapping errors associated with CLEAN, and there is no reason to 
believe these should be much smaller than those for $Q$.  The results of the 
fitting do not depend critically on the specific value of this weighting 
factor. The values of $\sigma_I$ and $\sigma_Q$ are given by the formulas 
recently proposed by Hovatta et al. (2012), which are based on Monte Carlo
simulations, and include contributions associated with thermal noise and with 
uncertainty introduced by the CLEAN process; the contribution of residual D-term 
uncertainty to $\sigma_Q$ is negligible for our case, far from the total 
intensity peak. The thermal-noise component was determined in regions far from 
the region of source emission.

As  the number density of electrons and the magnetic field strength along 
the slice are unknown, the database profiles were scaled so that the maximum 
total intensities of the observed and model profiles are equal. Matching the 
total flux densities of the observed and calculated $I$ profiles 
would be more accurate; however, doing this would drastically increase the 
computational time required to complete the comparisons, as each theoretical 
profile would have to be numerically integrated. In addition, convolution 
with a beam to mimic the effects of finite resolution removes most of the 
asymmetries in the theoretical $I$ profiles. As a result, most of the observed 
$I$ profiles are roughly Gaussian in shape. Thus, the scaling factors 
corresponding to matching the observed and theoretical profile maxima or
the observed and theoretical total fluxes are very similar, justifying our
approach. 

It is not possible to reliably apply standard statistical approaches to 
estimating the uncertainties in the resulting parameters without having well 
determined uncertainties in the fitted quantities --- $\sigma_I$ and $\sigma_Q$.
Unfortunately, estimation of the uncertainties associated with 
values in individual pixels of the $I$ and $Q$ images is not straightforward, 
since the imaging process is complex and the values in neighbouring pixels will 
be correlated, due to convolution with the CLEAN beam. 
Our $\sigma_I$ and $\sigma_Q$ estimates should typically correspond to
uncertainties in individual pixels that 
are correct to within a factor of a few; nevertheless, these estimates may not
be good enough to yield fully accurate $\chi^2$ values. Therefore, although the
calculated $\chi^2$ values can certainly be used to compare the profiles
correspond to
yielded by different sets of model parameters and identify a set of parameters
yielding a best fit, the $\chi^2$ values cannot be used to evaluate the overall
goodness of the fits obtained; an alternative method used to obtain
estimates of the uncertainties in the fitted parameters in Section~5
is described below. 

This method was applied to profiles for slices along which the EVPA was 
either parallel or perpendicular to the jet direction.  The Stokes parameters 
$Q$ and $U$ are defined such that $Q$ is positive and $U=0$ when the EVPA 
is parallel to the jet.

\section{Markarian 501 (Mrk~501)}

\begin{table*}
\centering
\caption{Best fit Parameters for Mrk~501 Slice~2 at 3 different wavelengths}
\begin{tabular}{@{}lccccc}
\hline
Wavelength & $\gamma^{\prime}$ & $\delta^{\prime}$ & $f$ & Epoch & Intrinsic Jet Width\\
\hline
4 cm & $53^{\circ} \pm 1^{\circ}$ & $83^{\circ} \pm 2^{\circ}$ & $0.40\pm 0.20$ & February 1997 & 5.5 mas\\
6 cm & $53^{\circ} \pm 1^{\circ}$ & $80^{\circ} \pm 2^{\circ}$ & $0.40\pm 0.05$ & February 1997 & 5.7 mas\\
13 cm & $54^{\circ}\pm 1^{\circ}$ & $86^{\circ} \pm 2^{\circ}$ & $0.35\pm 0.05$ & May 1998 &  21.6 mas  \\
18 cm & $54^{\circ}\pm 1^{\circ}$ & $81^{\circ} \pm 2^{\circ}$ & $0.60\pm 0.05$ & May 1998 &  22.9 mas  \\
\hline
\end{tabular}
\end{table*}

\begin{figure}
\centering
\includegraphics[width=0.5\textwidth]{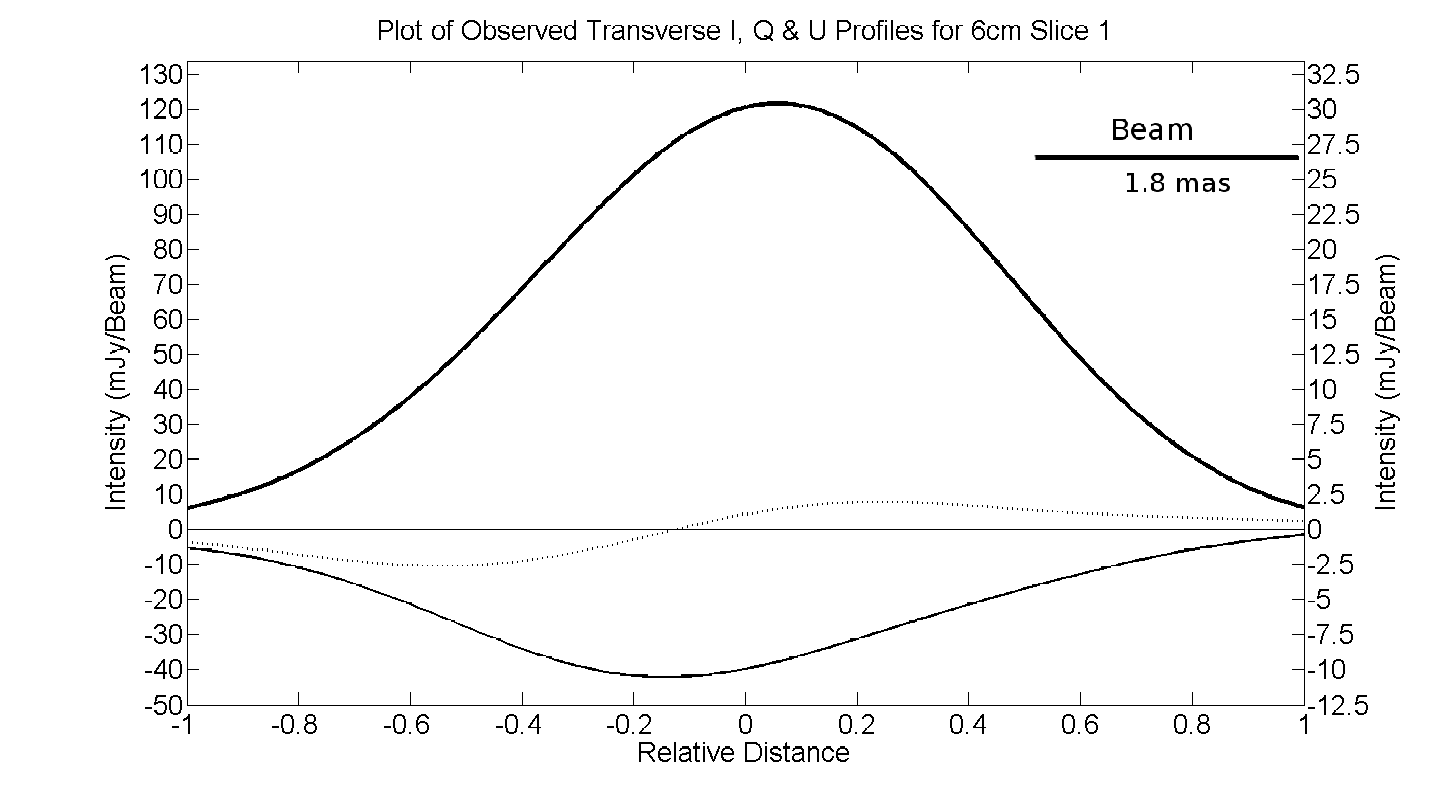}
\includegraphics[width=0.5\textwidth]{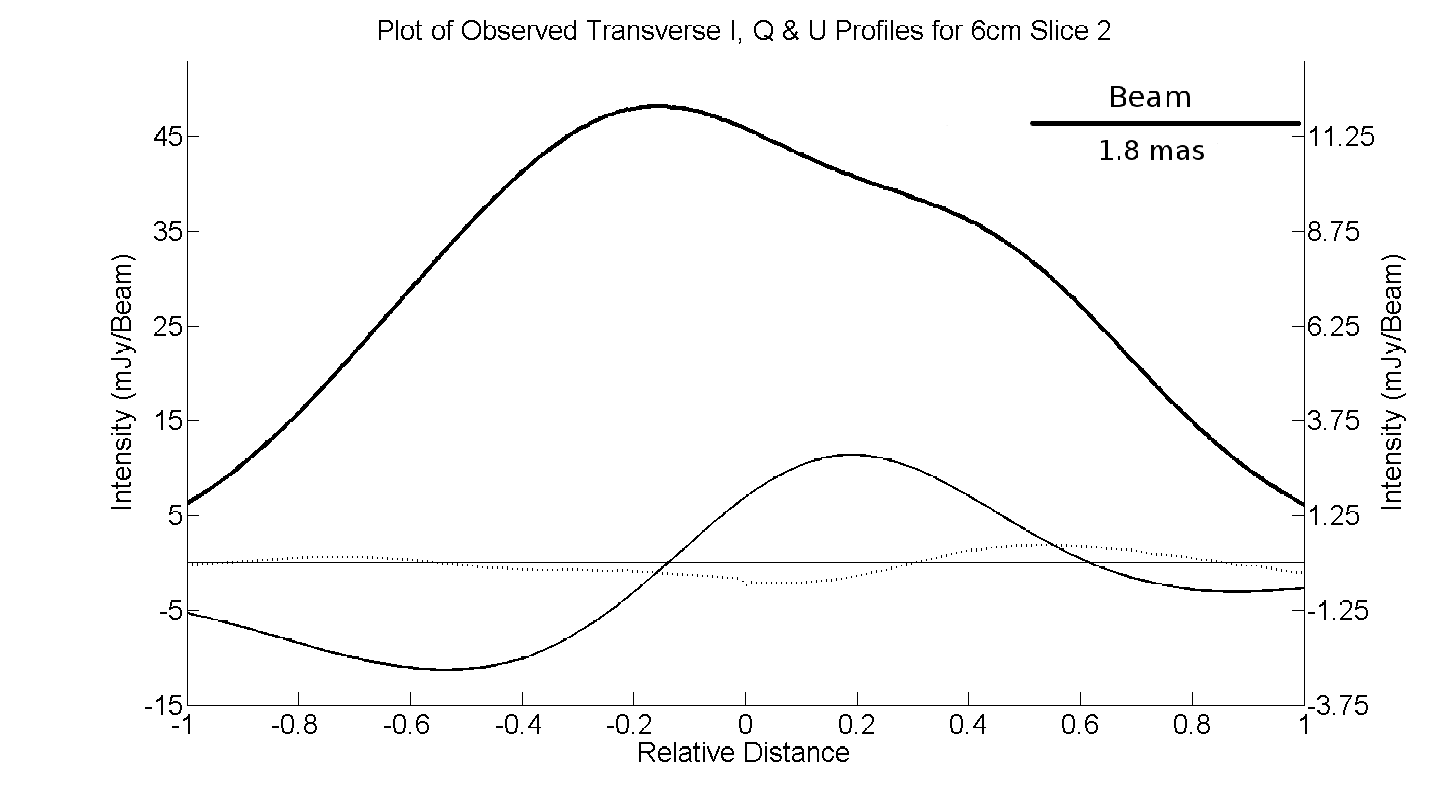}
\includegraphics[width=0.5\textwidth]{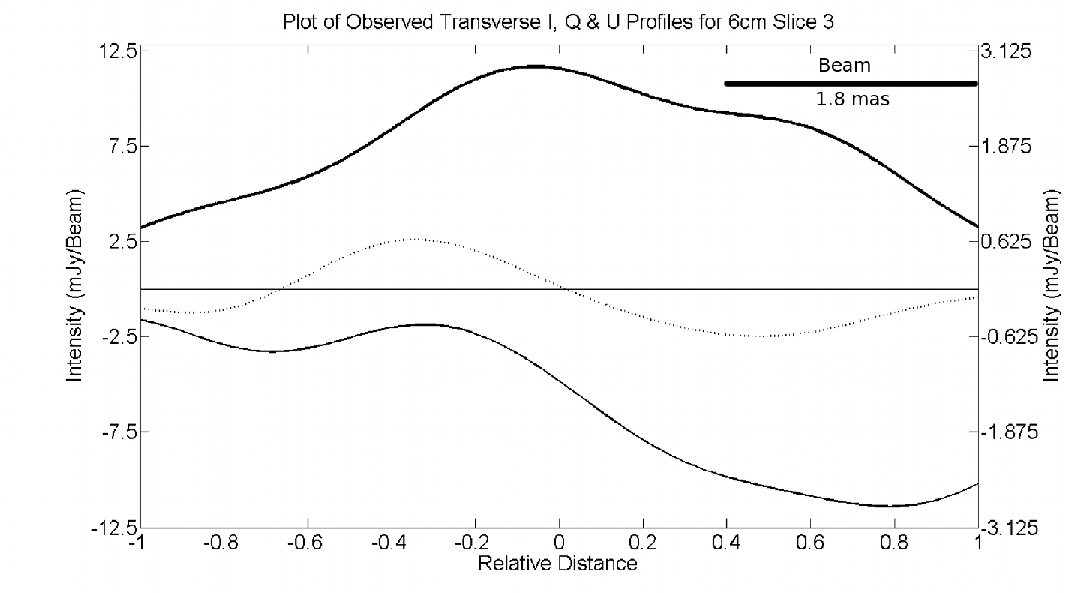}
\caption{Observed $I$ (solid), $Q$ (dashed) and $U$ (dotted) 
profiles for Slices 1, 2 and 3 (top to bottom) of the 6~cm 
Mrk501 map (see Fig.~3). The scale for the $I$ profile is given along the left-hand
vertical axis and the scale for the $Q$ and $U$ profiles along the 
right-hand vertical axis. The beam is 1.8~mas along the slice and 2.4~mas
transverse to the slice.}
\label{fig:Slices_iqu_6cm}
\end{figure}

\begin{figure}
\centering
\includegraphics[width=0.5\textwidth]{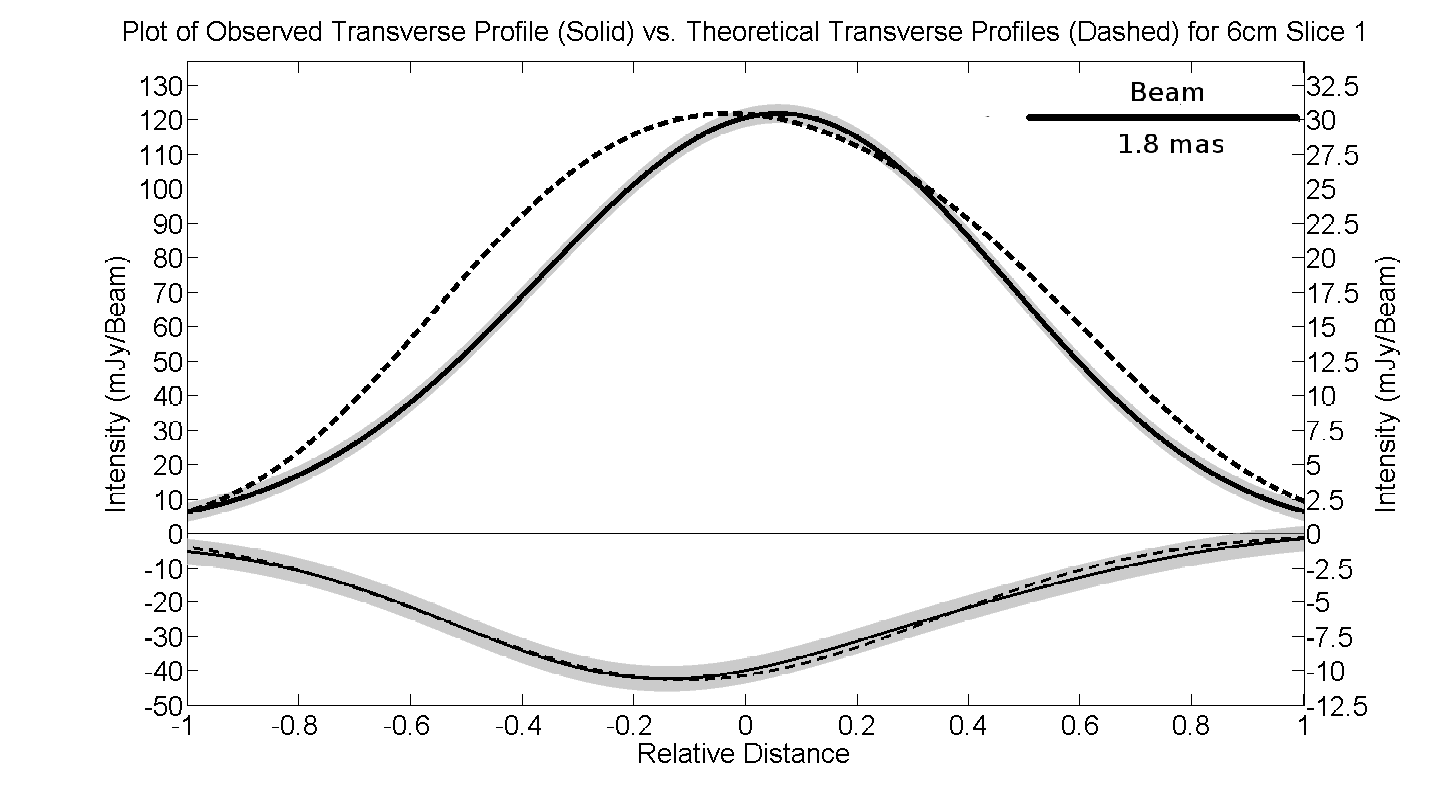}
\includegraphics[width=0.5\textwidth]{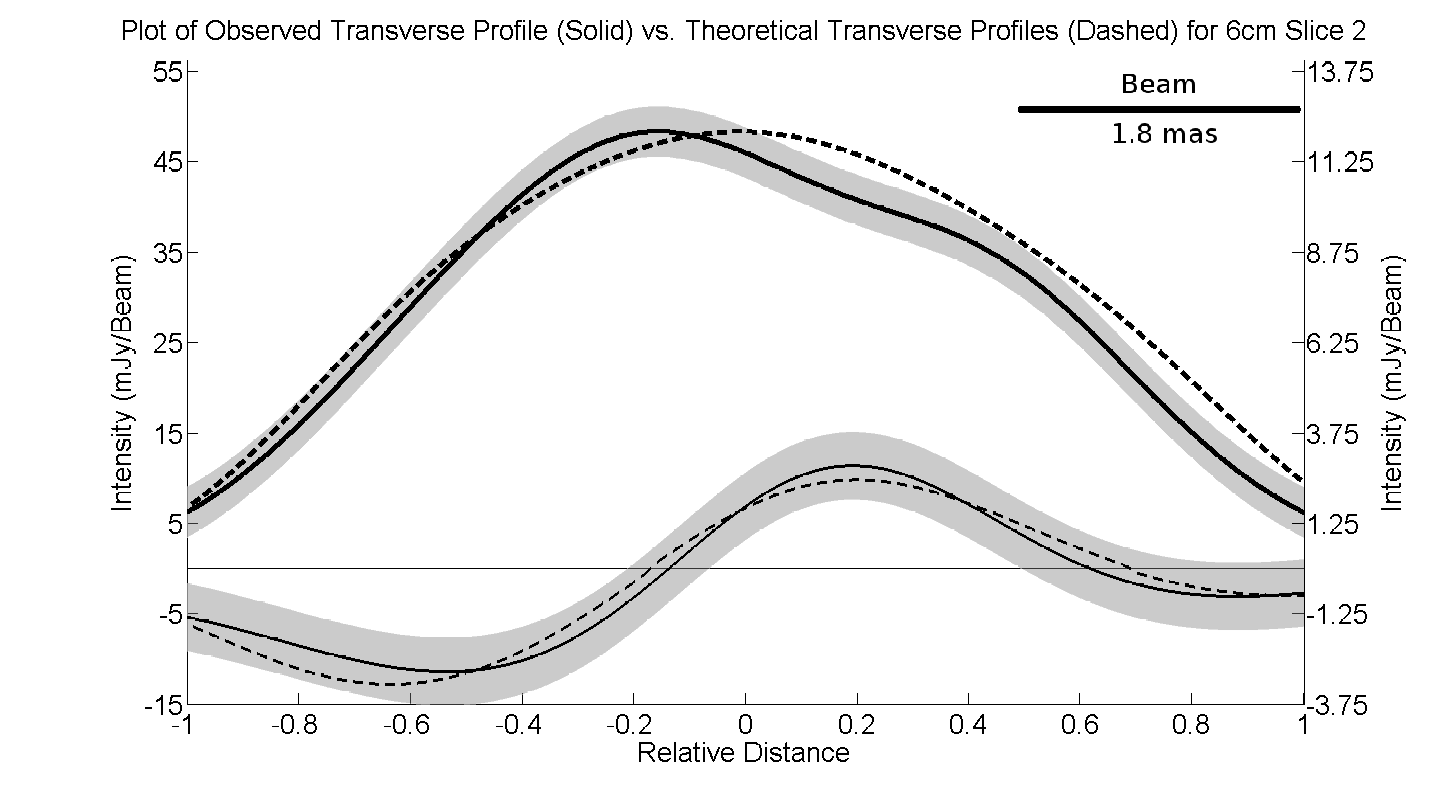}
\includegraphics[width=0.5\textwidth]{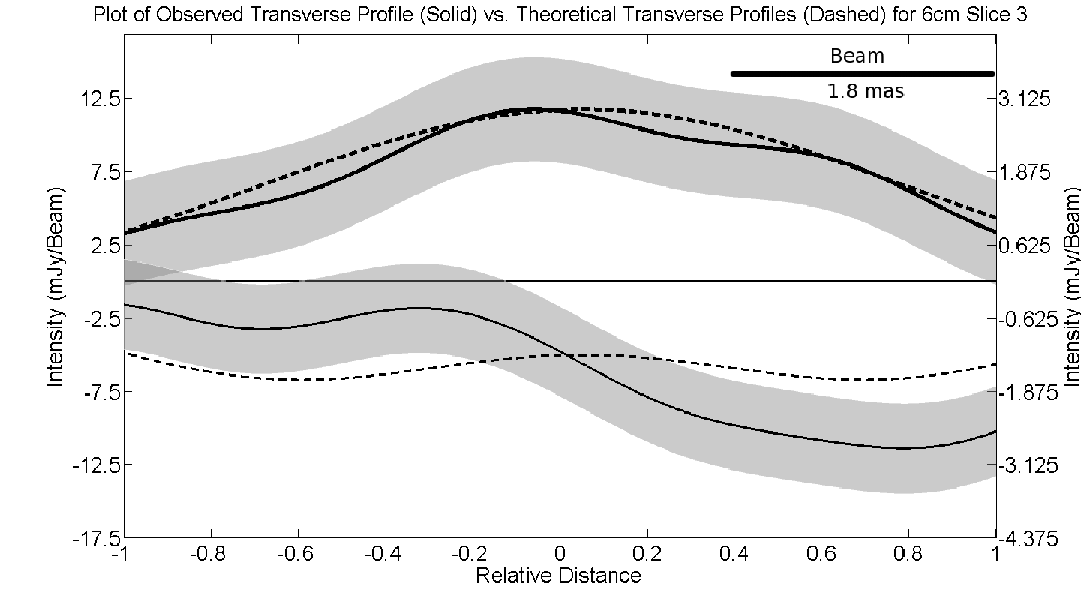}
\caption{Plots of observed and best-fit model $I$ and $Q$ profiles 
for Slices 1, 2 and 3 (top to bottom) of the 6~cm Mrk501 map (see Fig.~3). The 
observed profiles are solid and the best-fit profiles dashed; the
gray shaded areas surrounding the observed profiles correspond to
the range of the $1\sigma$ uncertainties for $Q$ and the $3\sigma$
uncertainties for $I$. The scale for the $I$ profile is given along the left-hand
vertical axis and the scale for the $Q$ profile along the 
right-hand vertical axis. The beam is 1.8~mas along the slice and 2.4~mas 
transverse to the slice.}
\label{fig:Slices_iqfit_6cm}
\end{figure}

\begin{figure}
\centering
\includegraphics[width=0.5\textwidth]{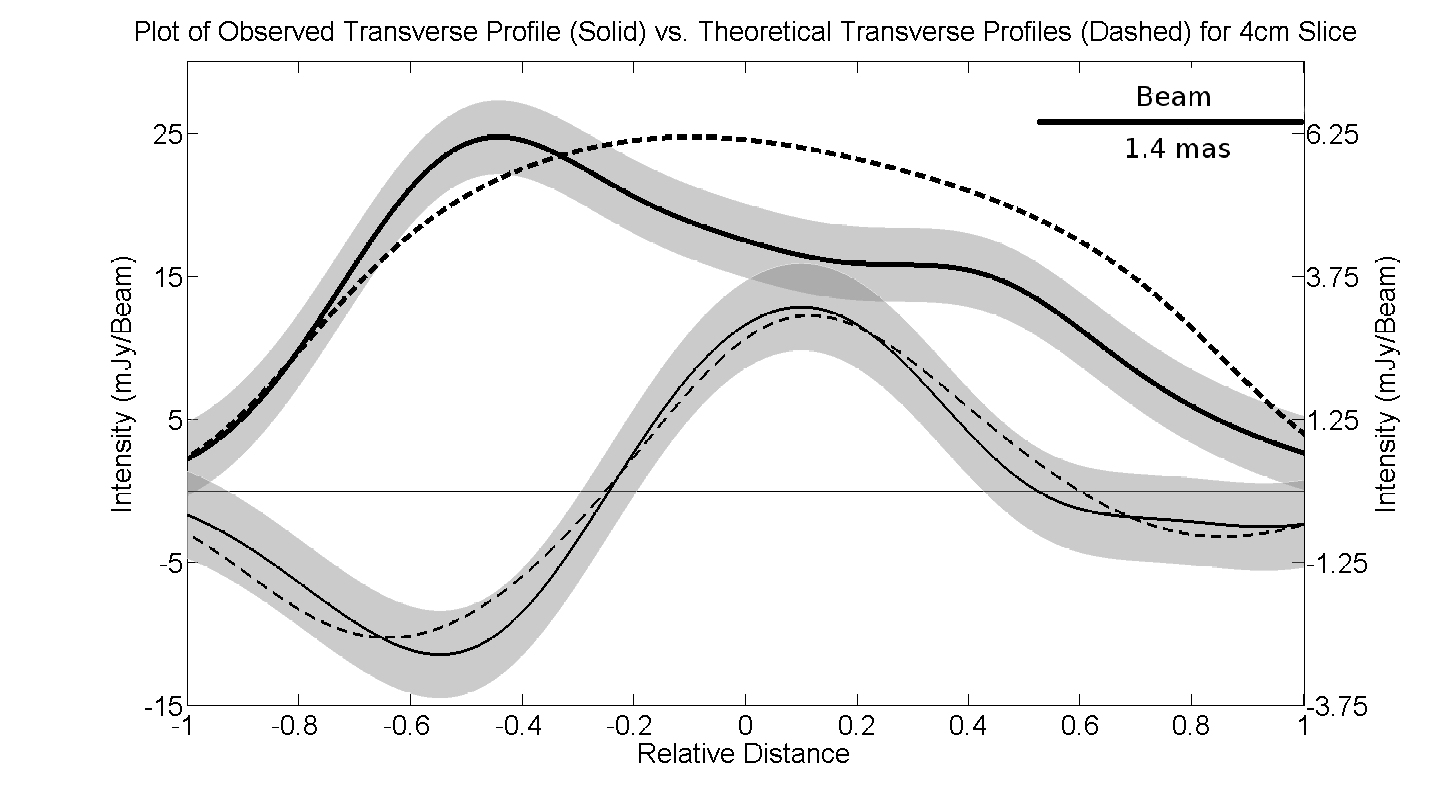}
\includegraphics[width=0.5\textwidth]{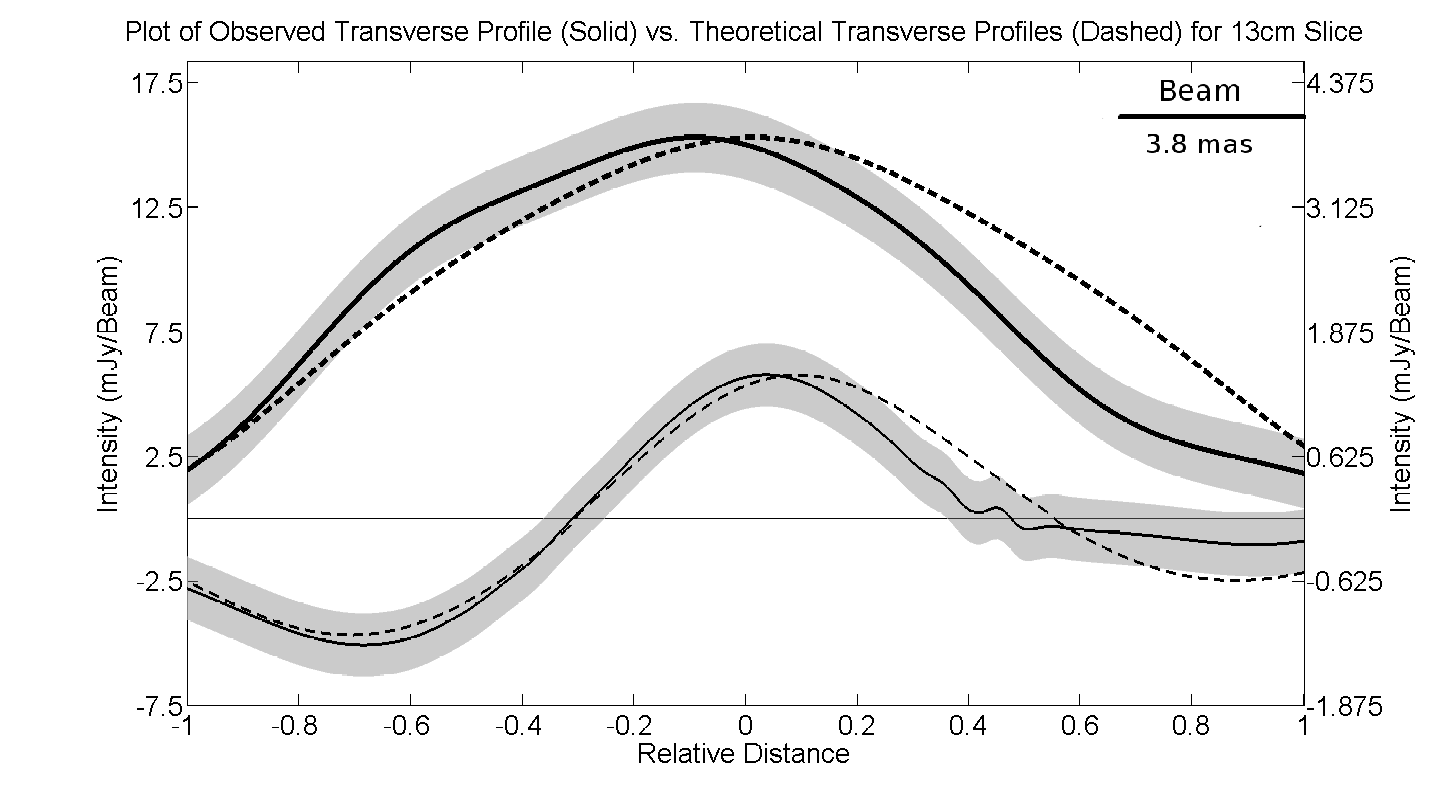}
\includegraphics[width=0.5\textwidth]{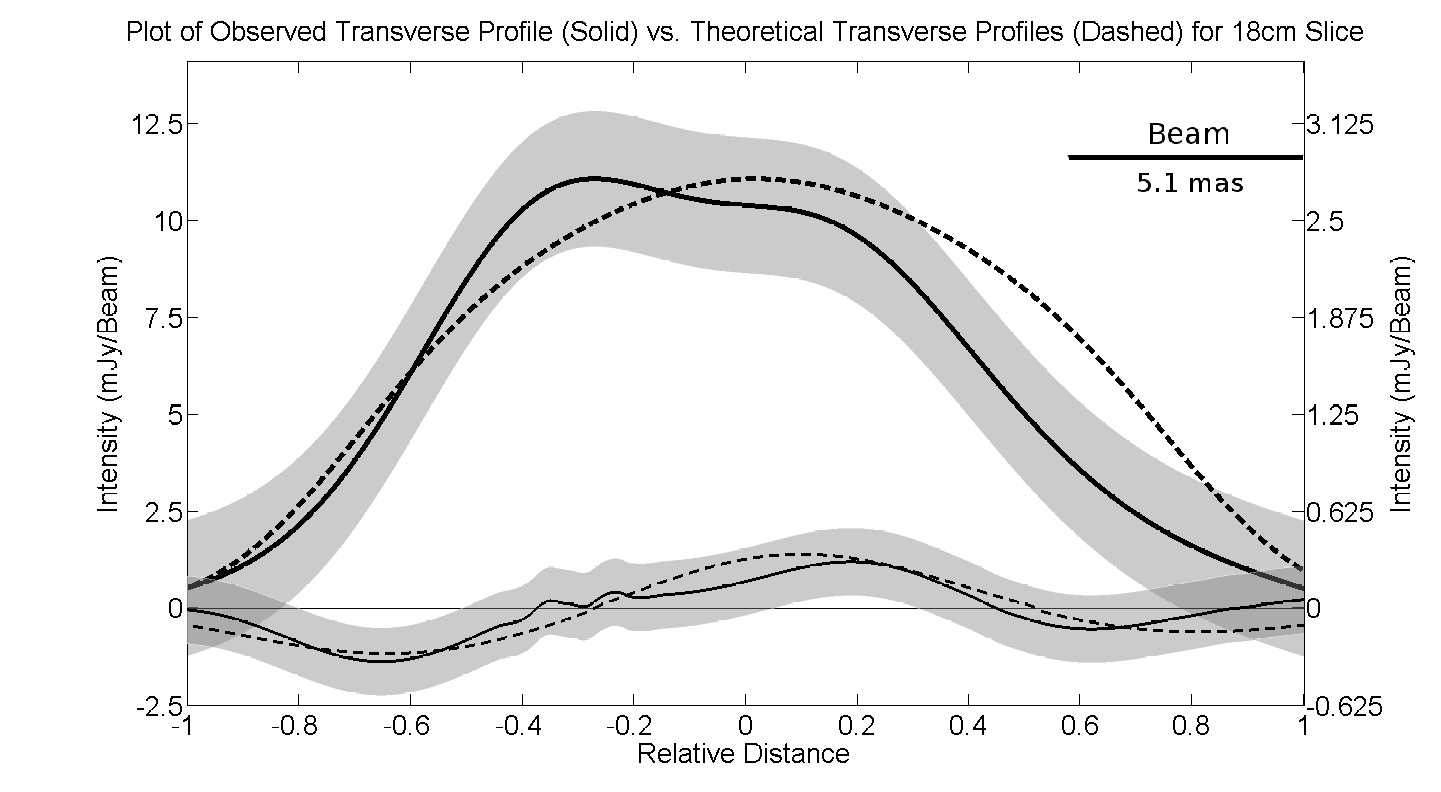}
\caption{Plots of observed and best-fit model $I$ and $Q$ profiles 
for 4~cm (top), 13~cm (middle) and 18~cm (bottom) slices taken in the
same region as Slice 2 (see Fig.~3).  The observed profiles are solid and 
the best-fit profiles dashed; the gray shaded areas surrounding the 
observed profiles correspond to the range of the $1\sigma$ uncertainties 
for $Q$ and the $3\sigma$ uncertainties for $I$. The scale for the $I$ 
profile is given along the left-hand vertical axis and the scale for the 
$Q$ profile along the right-hand vertical axis. The beam sizes along and
transverse to the slice are 1.4~mas and 1.6~mas (4~cm), 3.8~mas and 2.8~mas
(13~cm) and 5.1~mas and 3.8~mas (18~cm), respectively.}
\label{fig:Slices_4+13+18}
\end{figure}

An ideal VLBI jet for an analysis based on the method described in 
Section~4 would be one that is straight, well resolved and shows clearly 
visible transverse $I$ and $Q$ structure, with $U$ small compared to $Q$. 
Well-resolved VLBI jets are rare, however, and most VLBI jets contain some 
bends (though many of these are most likely small bends amplified by projection). 
Thus, VLBI jets displaying clear transverse
structure, especially in polarization, were sought, with the aim of determining 
whether profiles across such jets could plausibly be represented using
the helical field model outlined in Section 2. This requires measuring
these profiles away from bends, along directions orthogonal to the local jet
direction.

The well known active galaxy Mrk~501 (1652+398), which has been classified 
as a BL~Lac object with redshift $0.034$  \citep{Vaucouleurs}, has been 
chosen for this initial study.  The jet 
of Mrk~501 is almost certainly relativistic, as is shown by its one-sidedness; 
however, the shapes of the observed jet intensity and polarization profiles 
should be unaffected by this
relativistic motion. Piner et al. (2010) have found apparent component speeds
significantly less than the speed of light in the Mrk~501 jet. At first sight,
this would suggest a non-relativistic jet, at variance with the
one-sidedness of the jet (unless the VLBI jet were intrinsically one-sided).
In the context of the standard model for VLBI jets, it seems far more
likely that these low component speeds represent either relativistic motion
at a very small angle to the line of sight or pattern speeds that do not
directly reflect the speeds of the emitting plasma. In fact, a possible 
detection of superluminal 
motion with $v = (3.3\pm 0.3)c$ based on 43 and 86-GHz VLBI images has 
been reported  for this source (Piner et al. 2009). The most detailed
multi-wavelength studies of Mrk~501 have been carried out by Giroletti
et al. (2004, 2008). They derived constraints on the intrinsic speed of the
jet $\beta$ and angle of the jet to the line of sight in the observer's
frame $\delta$ on various scales, finding $\delta \leq 27^{\circ}$ and
$\beta \geq 0.88$, but with high values $\beta > 0.95 $ allowed only for
$10^{\circ}< \delta < 27^{\circ}$. They also found evidence for limb
brightening in total intensity on somewhat larger scales than those studied
here, which they attributed to transverse velocity
structure of the jet. In particular, they proposed that the jet has a 
a fast spine and a slower sheath, as suggested by Laing (1996), 
with the two experiencing different degrees of Doppler boosting. The observed
limb brightening could also have other origins, such as an enhancement in
the synchrotron emission coefficient at the edges of the jet due to
interaction of outer layers of the jet with the surrounding medium, or
a helical magnetic field confined to a thin shell (e.g. Laing 1981).

Our selection of Mrk~501 for our analyses is motivated by the fact that 
VLBI images of this AGN show a variety of transverse structures that could 
potentially be associated with a helical jet magnetic field, most
notably a fairly clear `spine--sheath' polarization structure, corresponding
to configuration 3 of Fig.\,2 (e.g. Pushkarev et al. 2005).  In addition,
being at a relatively low redshift, the VLBI jet of Mrk~501 is relatively
well resolved, compared to VLBI jets in more distant AGNs. 

Fig.\,3 shows a 6~cm total intensity map of this source with the polarization 
position-angle sticks superimposed, constructed from the same data as those of 
Pushkarev et al. (2005), from February 1997.  Near the core, the EVPAs 
are predominantly perpendicular to the jet  direction,  maintaining
this orientation as the jet direction begins to change about 6\,mas from
the core. Beyond this region, the EVPAs are perpendicular to the jet near 
the two edges and parallel in the centre, forming the spine--sheath 
polarization structure referred to above. 
The observational results of Giroletti et al. (2004, 2008) on
comparable scales to those studied here are broadly
similar to those presented in Fig.~3, though at higher resolution, so that
the transverse asymmetry is more apparent.

The choice of distances along the jet at which to analyse the transverse profiles 
is constrained by the need to find places where the EVPAs are either parallel 
or perpendicular to the jet (i.e., where $U$ is close to close to zero all 
across the jet), and the $Q$ profiles have sufficiently high 
signal-to-noise. The profiles to be analyzed
were constructed at locations where these conditions are satisfied, 
far from positions where the jet bends.
A further constraint on choice of position arises due to the fact that the 
jet is essentially unresolved in the immediate vicinity of the core. 

The three 6~cm slices chosen for analysis are marked on Fig.\,3.  The profiles 
were sampled using the 
AIPS task `SLICE', then compared to the database of 
model profiles (each convolved with the Gaussian observing beam) as described 
in Section 4,  and the values of $\gamma^{\prime}$, $\delta^{\prime}$ and 
$f$ resulting in the best fit (minimum residual) identified (given in 
Table 1). The intervals between adjacent 
values of $\gamma^{\prime}$, $\delta^{\prime}$ and $f$ in the
database were $1^{\circ}$, $1^{\circ}$ and $0.05$, respectively. 

The observed $I$ (solid), $Q$ (dashed) and $U$ (dotted) profiles for 
the three 6~cm slices are shown in Fig.\,4.  The left and right sides of the 
horizontal axis in this figure correspond to the North and South sides of 
the Mrk~501 jet.  Recall 
that the Stokes parameters are defined such that $Q$ is positive and $U$ is zero 
for a polarization $E$ field parallel to the jet. $U$ must be small (much less 
than $Q$) for the model used to be valid; the plots show that this condition is 
satisfied for the three slices chosen. 

The observed (solid) and best-fit model (dashed) profiles $I$ and $Q$ for 
the three slices are shown in Fig.\,5.  Here also, the left and right sides 
of the horizontal axis correspond to the North and South sides of the Mrk~501 
jet.  The $I$ and $Q$ curves in Fig.\,5 are easily distinguishable, 
as the $Q$ curves are much smaller in amplitude; the intensity scale for 
$I$ is shown to the left and that for $Q$ to the right.  
The shaded bands around
the $I$ and $Q$ profiles in Fig.\,5 correspond to $3\sigma$ and $1\sigma$ 
uncertainties respectively, estimated in accordance with the approach of 
Hovatta et al. (2012). Unconstrained fits for slice 3 yielded best-fit 
parameters corresponding to an opposite sense of the helicity of the magnetic
field, compared to slices 1 and 2; this corresponds to the fact that $Q$
changes from negative to positive across slice 2, but becomes more 
negative across slice 3 (Fig.~4). Since such a change in helicity between
slices 2 and 3 is physically implausible, we obtained fits to slice 3
constraining the sense of the helicity of the magnetic field to be the same
as for slices 1 and 2; this is the fit shown in Fig.~5 (and given in Table~1). 
The best-fit model $Q$ profiles for slices 1 and 2 lie 
within $1\sigma$ of the observed $Q$ profile, while the best-fit $Q$ profile
for slice 3 lies within $2\sigma$ of the observed profile. The model 
and observed $I$ profiles for slices 1  and 2 differ by more than $3\sigma$
over sustantial fractions of these profiles, with the model profile being 
more symmetrical than the observed profile; the fitted and observed
$I$ profiles for slice 3 agree to within about $2\sigma$. 

Since the intrinsic (or unconvolved) angular jet width was estimated using the 
procedure outlined in Section 4, the fitting procedure was repeated for a series 
of intrinsic jet widths centered on the values used above and listed in Table 1. 
Changing the intrinsic jet width had only a very minor effect on the best fit 
$\gamma^{\prime}$ and $\delta^{\prime}$ values while having a more pronounced 
effect on the degree of entanglement.  However, in general, as the intrinsic jet 
width varied from its estimated value, the $\chi^2$ value of the best fit profiles 
increased.  While the intrinsic jet width that minimised the $\chi^2$ value for 
a given slice is not exactly that found by fitting the $I$ profile, the two 
never differed by more then $10\%$. Values differing by more than $15\%$ yielded
very large values of $\chi^2$. 

As was noted above, the uncertainties in the fitted parameters cannot
be estimated reliably using standard statistical techniques, since 
the values of $\sigma_I$ and $\sigma_Q$ used are only estimated 
uncertainties.  Instead, estimates of the parameter uncertainties were
derived by carrying out Monte Carlo simulations, as follows.  Infinite 
resolution model maps corresponding to model parameters corresponding to those 
inferred for the Mrk501 jet were generated, and the Fourier transforms of 
these maps sampled at the set of baselines corresponding to the observations.  
This effectively yielded model visibilities at the set of UV baselines that 
were actually observed. Noise consistent with the noise levels observed in 
the images was then added. The resulting `noisy' model data were imaged 
in AIPS in the usual way, profiles were taken across the resulting maps and 
these profiles were subject to the same fitting process as the real observed 
profiles. The resulting best-fit parameters were then compared to the 
parameters used to generate the model maps to estimate the uncertainty 
in the fitting process.  This was carried out for 200 Monte Carlo realizations 
of the model data, and the distribution of derived uncertainties used to 
estimate the corresponding $1\sigma$ uncertainties, listed in Tables 1
and 2. 

The line-of-sight angles, $\delta^{\prime}$, for the first two slices in Table~1
agree to within a degree, and differ by less than $1\sigma$, as expected if the 
intrinsic bends in this region of the jet are small.  The value of 
$\delta^{\prime}$
for slice 3 differs by about $10^{\circ}$ ($\delta^{\prime} = 
90^{\circ}$), which corresponds to about $3.5\sigma$; however,
as was pointed out above, 
it was necessary to constrain the fit for slice 3 to have the same 
sense of helicity as slices 1 and 2. Therefore, we do not feel 
confident that the change in viewing angle implied by the nominal 
(constrained) best fit for slice 3 is significant.  

The fitting results 
indicate a somewhat higher value of $\gamma^{\prime}$ for slice 2 
than for slice 1, with this difference being about $4\sigma$, suggesting 
that the variation in $\gamma^{\prime}$ between
the slices may be real.  In this case, this suggests that 
the appearance of a spine--sheath polarization structure at the position 
of slice 2 could be due to a (relatively small) change in the 
helical pitch angle. The value of $\gamma^{\prime}$ for slice 3 lies between
the values for the other two slices; formally, $\gamma^{\prime}$ for slice
3 differs from the value for slice 2 by about $3.5\sigma$, and from the
value for slice 1 by slightly less than $2\sigma$. 

The values of $f$ appear to decline with distance from 
the core, from $0.7$ at slice 1 to $0.4$ at slice 2. This trend may continue
for slice 3 ($f = 0$), although this is unclear due to the uncertainty 
associated with slice 3 discussed
above.  The difference between slices 1 and 2 appears to be 
significant: the two $f$ values differ by approximately $4\sigma$; physically, 
this represents a decrease in the disordered component of the magnetic 
field with distance from the core in the region of the jet sampled.

The fitting procedure was repeated for slice 2 using images at wavelengths of
4\,cm for February 1997 (data of Pushkarev et al. 2005) and $13$\,cm 
and 18\,cm for May 1998 (data of Croke et al. 2010). The results are given in 
Table 2 and shown in Fig.~6. It is clear that the observations at 
these other wavelengths yield 
consistent best-fit parameter values; the values of $\gamma^{\prime}$, 
$\delta^{\prime}$ and $f$ essentially all agree to within $1\sigma$,
with the largest differences not exceeding about $2\sigma$. 

Fits were also obtained for Slices 1, 2 and 3 using Model B of Laing 
(1981), which consists of a cylinder of fixed radius containing a randomly 
orientated magnetic 
field with no radial component. The only parameter in this model is the viewing 
angle. This model can produce a region of transverse field at the jet axis 
surrounded by regions of longitudinal field, but the profiles produced by this 
model are completely symmetric.  Because Model B also produced profiles that are 
considerably more strongly polarized than is observed, we included the degree of
entanglement $f$ in these models as well.  Figure~7 shows the best-fit profiles 
for the three slices for the 6cm image of Mrk501. The model fits for slices
1 and 2 are clearly
considerably poorer than those in Fig.~5.  In contrast to Fig.~5, where the 
model and observed $Q$ profiles agree to within $1\sigma$ everywhere, the 
differences between the model and observed $Q$ profiles in Fig.~7 exceed 
$1\sigma$ essentially everywhere, with the differences exceeding $3\sigma$ 
along most of the profiles. The fit for Model B for slice 3 is very similar
to the corresponding fit for the helical-field model in Fig.~5. Thus, while
Model B of Laing (1981) can provide a comparably good fit for slice 3, it
does not provide a viable alternative to the polarization structure across
the other two slices.

\section{Spectral Index}

\begin{figure}
\centering
\includegraphics[width=0.5\textwidth]{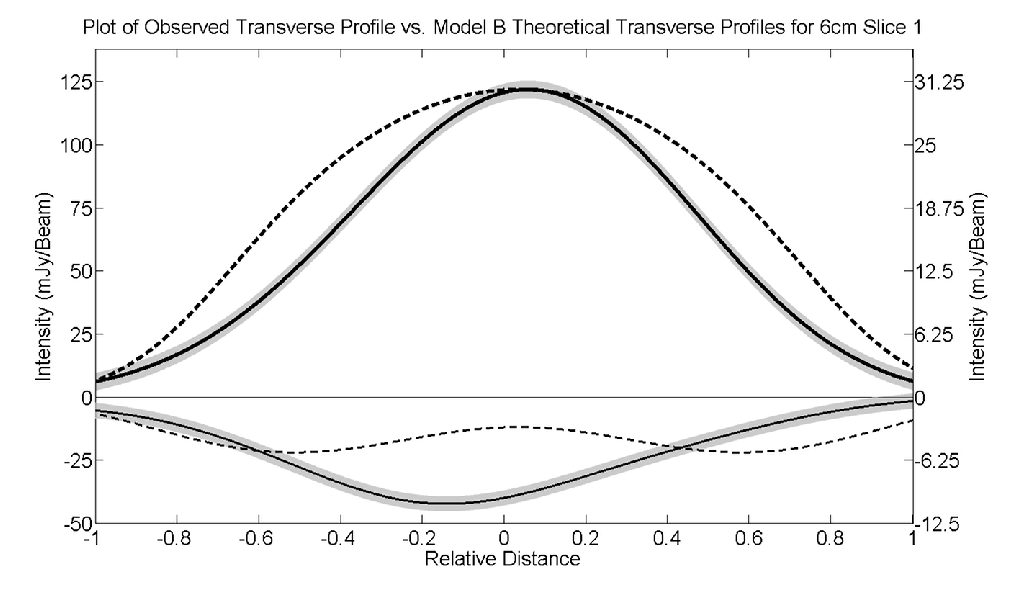} 
\includegraphics[width=0.5\textwidth]{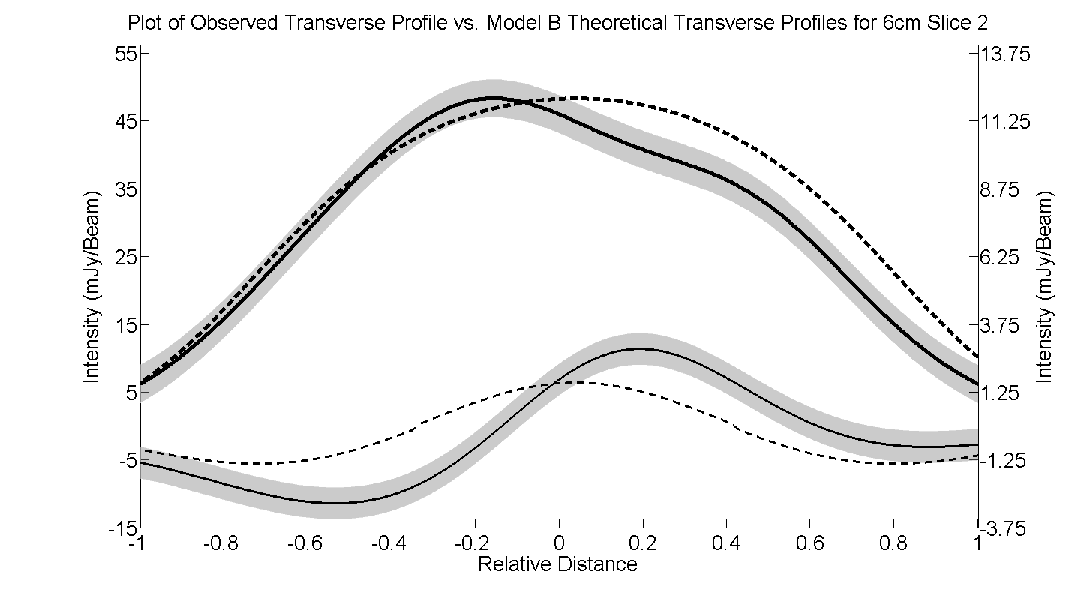} 
\includegraphics[width=0.5\textwidth]{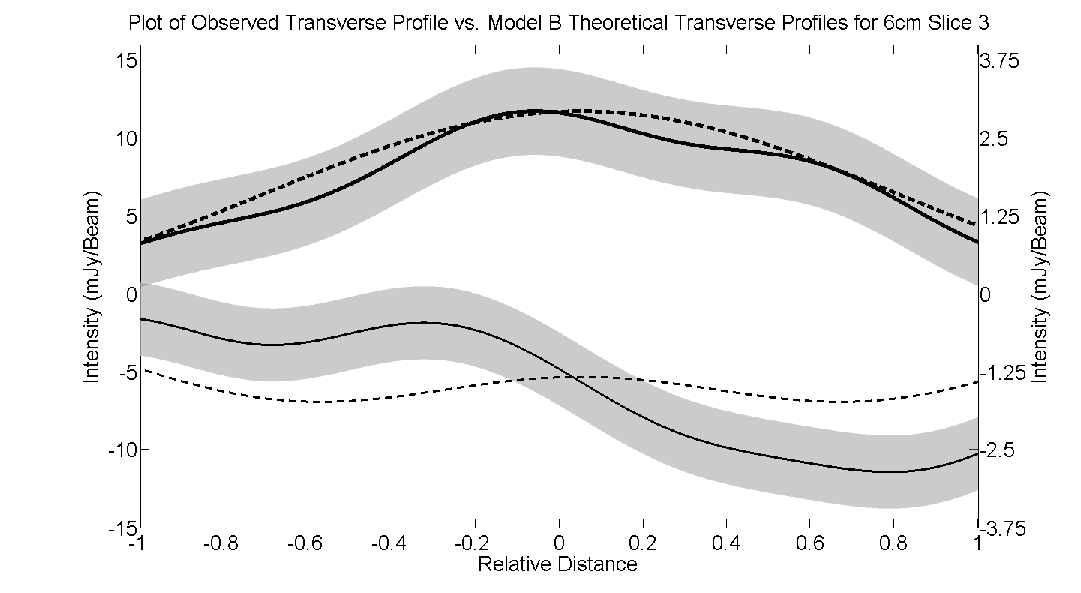} 
\caption{Plots of observed and best-fit model $I$ and $Q$ profiles 
for 6~cm slices 1, 2 and 3 obtained for Model B of Laing (1981).  
The observed profiles are solid and the best-fit profiles dashed; the gray 
shaded areas surrounding the observed profiles correspond to the range of the 
$1\sigma$ uncertainties for $Q$ and the $3\sigma$ uncertainties for $I$. The 
scale for the $I$ profile is given along the left-hand vertical axis and the 
scale for the $Q$ profile along the right-hand vertical axis. 
The beam is 1.8~mas along the slice and 2.4~mas 
transverse to the slice.}
\label{fig:ModelB}
\end{figure}

\begin{figure*}
\centering
\includegraphics[width=1\textwidth]{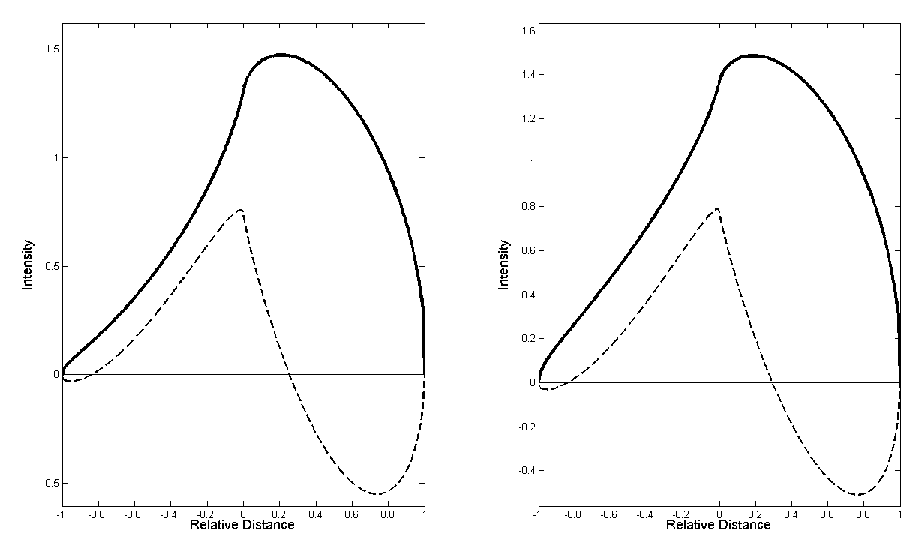}
\caption{Theoretical profiles generated numerically for $\gamma^{\prime}=60\degree$, 
$\delta^{\prime}=40\degree$ and $f=0$. The left and right profiles were determined 
for $\alpha=1.0$ and $\alpha=0.5$, respectively. The magnetic-field orientation is 
perpendicular to the polarization orientation.}
\label{fig:Spectral}
\end{figure*}

A spectral index ($\alpha$) of unity was assumed throughout the jet in order to 
make the equations that govern this model analytically solvable.
However, observations of the spectral indices in the jets of AGN have shown that 
this value is typically $\simeq 0.5$.  This gives rise to concerns about the 
trustworthiness of the fits obtained assuming $\alpha=1.0$. To investigate this, 
model profiles were obtained numerically for two other spectral indices. 

Profiles for both $I$ and $Q$ were determined numerically by integrating equations 
A1 and A2 from Laing (1981) using the Gauss-Kronrod quadrature \citep{Kronrod}. 
This method was required as the functions governing both $I$ and $Q$ have 
singularities at one of their limits. 

Figure \ref{fig:Spectral}  shows the effect of variations in the spectral index 
on the $I$ and $Q$ profiles. The  left and right hand plots show profiles for 
jets with $\alpha=1$ and $\alpha=0.5$ respectively, and the same values of 
$\gamma$, $\delta$ and $f$. The differences between these profiles are very small. 
The positions where the polarization angle changes by 90$\degree$ are slightly 
different. In addition, the amount of longitudinal polarization has increased 
slightly, while the amount of transverse polarization has decreased. This second 
difference results in the configuration map for $\alpha=0.5$  being slightly 
different to that for $\alpha=1$ (compare Fig.~2 and 9). 

The effect of changing spectral index on the best fit model values of 
$\gamma^{\prime}$, $\delta^{\prime}$ and $f$ was examined using slices through 
the 13~cm images at the location of Slice 2 in Fig.~3. Model databases were then 
generated numerically using values of $\alpha = 0.507$ (the average value of 
$\alpha$ across the slice) and $\alpha=0.248$ (the minimum value of alpha across 
the slice, representing the maximum deviation from unity)  and the FWHM of the 
beam used in creating the 13~cm map.  These databases were then used to obtain 
new best-fit parameters for the 13~cm slice. The results are shown in Table 3. The 
 differences in the best fit parameters are extremely small with the 
$\gamma^{\prime}, \delta^{\prime}$ and $f$ values for the two numerically calculated 
best-fit profiles ($\alpha=.508$ and $\alpha=.248$) and the analytically calculated 
best-fit profile (for $\alpha=1.0$) all coinciding within the 1$\sigma$ errors.

\begin{table}
\centering
\caption{Best fit Parameters for 13~cm Mrk501 Slice for three values of $\alpha$}
\begin{tabular}{@{}lcccccc}
\hline
$\alpha$ & $\gamma^{\prime}$ & $\delta^{\prime}$ & $f$\\
\hline
1.000 & $54^{\circ}\pm 1^{\circ}$ & $86^{\circ}\pm 2^{\circ}$ & $0.35\pm 0.05$\\
0.507 & $53^{\circ}\pm 3^{\circ}$ & $85^{\circ}\pm 2^{\circ}$ & $0.40\pm 0.05$ \\
0.248 & $53^{\circ}\pm 1^{\circ}$ & $84^{\circ}\pm 2^{\circ}$ & $0.40\pm 0.05$ \\
\hline
\end{tabular}
\end{table}

\section{Comparison with Results from Faraday Rotation profiles}

As described earlier, the fractional polarization from the helical field model is 
generally lower on the side of the jet where the field lines are closest to 
the line of sight. If the magnetic field threading the Faraday rotating medium 
is comoving with the jet and is threaded by the same magnetic field, the magnitude 
of the Faraday rotation measure will  be highest on the side of the jet where the 
field lines are closest to the line of sight, i.e., on the side with the lower
fractional polarization. It is therefore of interest to check whether the 
transverse polarization and Faraday-rotation profiles observed 
for the Mrk~501 jet are consistent with this picture.  In particular, 
Gabuzda et al. (2004) reported the detection of a transverse 
Faraday rotation measure (RM) gradient across the Mrk~501 jet,
based on 2+4+6~cm VLBA polarization observations, and interpreted this as
evidence for a helical magnetic field associated with this jet. Croke et al.
(2010) subsequently reported a transverse RM gradient in the same sense based on 
4+6+13+18~cm VLBA polarization observations. 

Taylor \& Zavala (2010) have recently questioned the validity of reported
transverse RM gradients, and proposed a number of criteria for the
reliable detection of such gradients, namely that there be (i) at least 
three ``resolution elements'' (taken to mean three beamwidths) across the 
jet, (ii) a change in the RM of at least three times the typical error 
$\sigma$, (iii) an optically thin synchrotron spectrum at the location of 
the gradient, and (iv) a monotonic, smooth (within the errors) change in the 
RM from side to side. The more recent Monte Carlo simulations of Hovatta et 
al. (2012) suggest that a more reasonable limit for the distance spanned by the
RM values comprising the gradient is two beamwidths if the RM values differ
by at least $3\sigma$, and as little as 1.5 beamwidths in the case 
of higher signal-to-noise ratios.

As part of their study, Taylor \& Zavala (2010) consider several RM
maps based on multi-wavelength 2--4~cm VLBA data, searching for RM gradients
specifically in regions where the RM distributions span more than three
beamwidths. On this basis, they report an absence of significant transverse 
RM gradients in the Mrk~501 jet, based on profiles taken roughly 10~mas from
the core, and therefore call into question the earlier results of Gabuzda et 
al. (2004) and Croke et al. (2010).  We do not wish to analyze 
this discrepancy between the results of Gabuzda et al. (2004) and Croke et al. 
(2010), on the one hand, and Taylor \& Zavala (2010), on the other hand, in 
full detail here. However, for transparency, we will make some remarks concerning
this, as a justification for our further consideration of the results
of Gabuzda et al. (2004) below.

The transverse RM gradient reported by Gabuzda et al. (2004) essentially
satisfies the 
last three criteria of Taylor \& Zavala (2010), but not the first; however,
because the
gradient spans approximately 1.7 beamwidths, it marginally satisfies the 
modified criterion of Hovatta et al. (2012). The gradient reported by Croke 
et al. (2010) spans roughly four beamwidths across the jet, and satisfies 
all four criteria of Taylor \& Zavala (2010). Note that the profile 
considered by Taylor \& Zavala (2010) is further from the core than the region
considered by Gabuzda et al. (2004), where the magnitude of the RM values 
would be expected to be lower; the region considered by Croke et al. (2010) 
overlaps with the region of the profile of Taylor \& Zavala (2010), but the 
data of the former study were considerably more sensitive to small amounts 
of Faraday rotation, since they included observations at the longer 
wavelengths of 13 and 18~cm. 

Thus, while the resolution of the images of Gabuzda et al. (2004),
analyzed in the same general region as our own profiles, may not represent 
conclusive proof of a uniform transverse RM gradient, their results can 
reasonably be 
considered at least tentative evidence for a Faraday-rotation gradient across 
the Mrk~501 jet, particularly in light of the subsequent, better-resolved 
results of Croke et al. (2010). Therefore, we feel justified in using these 
results as a working hypothesis. We emphasize that we are interested here only 
in the very crude question of whether there seems to be a higher typical RM 
magnitude on one or the other side of the VLBI jet, and whether this is 
consistent with the side predicted by our simple model.
 
\begin{figure}
\centering
\includegraphics[width=1\columnwidth]{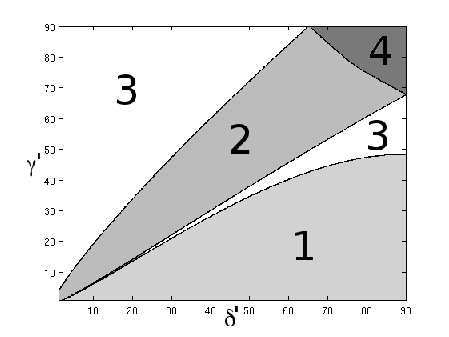}
\caption{Magnetic-field configuration map analogous to Fig.~2 for $\alpha=0.5$ 
after convolution with a Gaussian beam with a FWHM one-quarter the size of the 
profile to mimic the effects of finite resolution. The region numbers correspond 
to the configuration types listed in Section~3. }
\label{fig:SpectralConfig}
\end{figure}

\begin{figure}
\centering
\includegraphics[width=1.0\columnwidth]{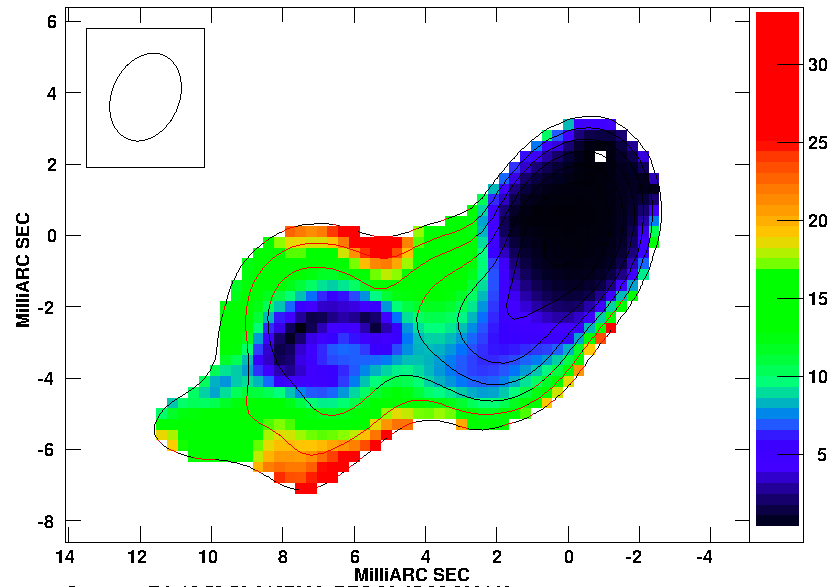}
\caption{6~cm total intensity contours given in Fig.~3 superposed with a colour
image of the distribution of the degree of polarization. The increase in the
degree of polarization toward the edges of the jet, also reported by Pushkarev
et al. (2005), is clearly visible. The locations of the highest degrees of polarization
do not coincide with those expected for curvature-induced polarization.}
\label{fig:mpol}
\end{figure}

In the regions of the jet considered here, comparing the two sides of the
jet, the Northern side (corresponding to the left-hand sides of the plots in 
Figs.\,5--6) generally displays higher fractional polarization, i.e., more 
negative values of $Q/I$.  It follows that the Faraday rotation measures on the 
Northern side of the jet, RM$_N$, are expected to be smaller in magnitude than 
those on the Southern  side, RM$_S$. This is consistent with the observational 
results of Gabuzda et al. (2004), based on the same 2--6~cm VLBA data considered
here: $\textrm{RM}_{N} = -55\pm 20$~rad/m$^2$, $\textrm{RM}_{S} = 130 \pm 20$~rad/m$^2$. The effect 
of the measured Galactic rotation measure in the direction toward Mrk~501
[$+42$~rad/m$^2$, Rusk (1988)] 
was subtracted from the observed polarization angles before the Faraday 
rotation 
map was constructed, so that the residual observed rotation measure should 
correspond to Faraday rotation occurring in the vicinity of the AGN. If the
Galactic rotation measure is instead taken to be the value typical of the region
within a few degrees of Mrk~501 in the catalog of Taylor, Stil and Sunstrum (2009), 
$\simeq +20$~rad/m$^2$, these two values become $\textrm{RM}_{N} \simeq -35$~rad/m$^2$, 
$\textrm{RM}_{S} \simeq 150$~rad/m$^2$. 

We can crudely estimate the expected quantitative difference between the magnitudes
of the Faraday rotation measures on either side of the jet if the Faraday rotating
material is concentrated in a relatively thin shell in outer layers of the jet
using Eq. (2) of Laing (1981): for $\gamma^{\prime} = 40^{\circ}$ and $\delta^{\prime} =
80$, this indicates a difference of about 35\%, somewhat smaller than the observed
value of about $60-75\%$.

Considering this one case, and given the limited transverse resolution of the
rotation-measure image of Gabuzda et al. (2004), it could be a coincidence that 
the model considered correctly predicts the side of the jet that should have 
the higher Faraday rotation. However, our analysis here illustrates
the sort of comparison that could, in principle, usefully be carried out once 
profile-fitting results for a greater number of AGNs are available. 
Of course, the sign of the Faraday rotation measures 
cannot be predicted by our model, as the polarization profiles do not depend on 
the polarity of the magnetic field. 

\section{Derivation of $\delta$ and $\beta = v/c$ in the Observer's Frame}

The line of sight angle $\delta'$ in the rest frame of the jet is related
to the corresponding angle in the observer's frame, $\delta$, by
\begin{equation}
 \sin \delta^{\prime}= \frac{\sin \delta}{\Gamma[1-\beta\cos\delta]} 
\end{equation}
where $\Gamma$ is the Lorentz factor for the motion and $\beta = v/c$ (Rindler
1990).

In addition, the apparent speed of a component moving along the jet 
can be written
\begin{eqnarray}
\beta_{app} &=& \frac{\beta \sin\delta}{1-\beta\cos\delta} \\
             &=& \Gamma\beta\sin\delta^{\prime} \\
             &=& (\beta^{-2}-1)^{-1/2}\sin\delta^{\prime} \\
\end{eqnarray}

It follows that, if $\beta_{app}$ is known from component motion and
$\delta^{\prime}$ is known, for example, from profile fitting such as
that carried out in this paper, Equation (8.4) can be rearranged to give
\begin{equation}
\beta = \left[ 1 + \frac{\sin^2\delta^{\prime}}{\beta_{app}^2}\right]^{-1/2}
\end{equation}

The profile fitting has 
indicated $\delta^{\prime} \simeq 83^{\circ}$ for the Mrk~501
jet. Taking this together with the superluminal motion reported by Piner et al. (2009), 
$\beta_{app} = 3.3$, gives $\beta \simeq 0.96$ and  $\delta \simeq 15^{\circ}$. 
Using the range of $\delta^{\prime}$ values given by the various profile fits,
$80^{\circ}-86^{\circ}$, does not change the resulting values of $\delta$ by
more than $1^{\circ}$.  These results are consistent with the conclusion
of Giroletti et al. (2004), based on completely different information, that 
$\delta \leq 27^{\circ}$ and $\beta \geq 0.88$, with $\beta > 0.95$ allowed only 
for $10^{\circ}< \delta < 27^{\circ}$.
Although these particular results are somewhat uncertain, 
since the relevant superluminal motion corresponds to a
weak feature whose position is difficult to determine precisely,
this illustrates the potential of this approach.  Note as well that
using the more typical subluminal speeds obtained for the jet of Mrk~501, e.g.
$\beta_{app} \simeq 0.47$ (Piner et al. 2010),
implies angles to the line of sight $\delta \simeq 60^{\circ}$ and $\beta \simeq
\beta_{app}$, which are not reasonable, since they cannot provide the
observed one-sided VLBI structure. This implies that the very low component
speeds typically observed in Mrk~501 are incompatible with its observed
polarization structure, and so must represent pattern speeds rather than 
physical speeds (in other words, they do not represent highly relativistic motion
viewed at a very small angle). 

The approach taken in this section is similar to that used by Canvin 
et al. (2005) on kiloparsec scales, in that both analyses fit a jet model to 
the observed profile in order to determine the angle between the jet axis and 
the line of sight in the rest frame of the flow. One advantage of applying this
technique on parsec scales is that it is possible to obtain constraints on the 
underlying flow speed fairly directly using observations of superluminal components,
whereas, in the kiloparsec case, more indirect estimate must be used, such as
the jet--counterjet brightness ratio. 

\section{Discussion}

In general, the procedure described above has yielded good fits to the 
polarization data for Mrk~501. The deviations between the model and observed 
$Q$ values for slices 1 and 2 are less than the estimated $1\sigma$ 
uncertainties for the 
observed $Q$ profiles,  while the discrepancies for the somewhat uncertain
slice 3 are less than $2\sigma$ (Hovatta et al. 2012). The model has 
thus successfully 
reproduced the main features of the observed polarization. The total 
intensity fits were much less successful, with the model profiles 
generally being 
appreciably more symmetrical; differences between the observed
and model $I$ profiles were within $2\sigma$ for the 6-cm Slice 3, but 
frequently exceed $3\sigma$ for the other slices considered. Whether  
this discrepancy between the model and observed $I$ profiles is due to other 
factors giving rise to asymmetry in the observed $I$ profiles or unsuitability 
of the model is not clear. For this reason, the results presented here should 
be considered tentative. 

Formally good model fits were not expected using the procedures described in 
this paper; the main goal was to determine whether a simple helical-field model 
such as that considered here is able to qualitative reproduce the shapes of the 
observed profiles, particularly in $Q$. In fact, there are many reasons why it
would be unreasonable to expect a formally good fit to both $I$ and $Q$.
The structure of the jet may differ in many ways from the simple structure used 
here.   For example, any deviation from perfect cylindrical symmetry, or an 
emissivity that changes with  distance from the jet axis, would introduce features 
that could not be reproduced by the model.  Some such physical deviations may be 
responsible for the double--peaked $I$  structure in the $3.6$\,cm profile in 
Fig.\,6, which is poorly represented by the model.  Attempting to introduce 
further parameters into the model would result in fits that were much more  
poorly constrained by the data. Given the simplicity of the model used, 
its ability to reproduce the $Q$ profiles so well is striking.

Thus, the most important result of our profile fitting is that even the simple 
helical-field model considered here can reproduce most of the qualitative 
features of the polarization profiles, lending support to the hypothesis
that a helical magnetic-field component is present in the jet of Mrk~501 on 
parsec scales.  

\subsection{Curvature-Induced Polarization?}

Given the observed bends in 
the jet of Mrk~501, it is natural to ask whether
the observed polarization 
structure is associated with curvature of the jet. 
As the jet bends, 
the jet plasma is squeezed and stretched at the inside and outside edges, 
respectively. If, as seems to be the case here, the magnetic field is 
highly disordered, but with a mean value of 
$<B_{long}^2>$ greater 
than $<B_{perp}^2>$, where $B_{long}$ and $B_{perp}$ are the magnetic fields
along and perpendicular to the jet, then squeezing the plasma along the jet
axis will increase $<B_{perp}^2>$, reducing the degree of 
polarization, while stretching the plasma in this same direction will have
the opposite effect.  This effect would increase/decrease the degree of 
polarization at the outside/inside edges of jet bends.

To investigate this, we constructed a map of the degree of polarization $m$ 
at 6~cm, shown in Fig.~10.  The observed fractional polarization is
greatest at the edges of the jet (reaching values $\simeq 25-30\%$), as is 
expected for curvature-induced polarization. 
However, the locations of highest fractional polarization are
on the {\em inside} edges of local bends in the jet (the northern
side of the jet between slices 1 and 2, and 
the southern side of the jet beyond slice 2), whereas curvature-induced
polarization should be highest on the {\em outside} edges of these bends.
Therefore, we find no evidence that the longitudinal polarization
observed at the edges of the Mrk~501 jet is primarily induced by the 
curvature of the jet as it bends.

\section{Conclusions}

A method for comparing model and data profiles for total intensity and polarization 
in astrophysical jets has been described and demonstrated. The method has been 
used to compare observed profiles of the parsec-scale jet of Mrk~501 with the 
predictions of a model in which a cylindrical jet is permeated by a magnetic 
field with a uniform helical component and a disordered component.

The jet of Mrk~501 shows several characteristics that are consistent with a 
helical jet magnetic field, most strikingly the spine--sheath transverse 
polarization structure observed in the region of our Slice 2.  The best model 
fits obtained correspond to 
pitch angles in the jet rest frame $\gamma^{\prime} \simeq 40-50^{\circ}$ and 
viewing angles in the jet rest frame $\delta^{\prime}\simeq 83^{\circ}$. 
These fits describe the polarization structure well, though the total 
intensity fits are generally poorer. However, given the extremely simple 
nature of the 
helical-field model, for example its rigid cylindrical symmetry and uniformity 
of emission, its success in reproducing the general features of the $Q$ 
profiles is noteworthy, and suggests that further comparisons, using higher 
resolution data and data for other sources, would be worthwhile. 

The best fit values of the line-of-sight angle are very similar for all the
analyzed 4~cm, 6~cm, 13~cm and 18~cm slices across the Mrk~501 jet, within the
estimated $1\sigma$ uncertainties (apart from the fitted viewing angle for
the somewhat more uncertain slice 3, which differs from the other values by
$2-3\sigma$). This suggests 
that, as expected, the large apparent changes in jet direction are in fact very 
small bends that are greatly amplified by projection. The fitted pitch angle
increases from the first to the second 6~cm slice, bringing about the observed 
transition in polarization structure (from configuration 1 to configuration 3 as 
described in Fig.\,2). The fraction of energy in the disordered 
magnetic-field component seems to decrease with distance along the jet. 

Together with the tentative superluminal speed reported by Piner et al. 
(2009), the estimate for the viewing angle in the jet rest frame obtained 
through the profile fitting, $\delta^{\prime}\simeq 83^{\circ}$, enables 
determination of the viewing angle and jet velocity in the obsever's frame, 
$\delta\approx 15^{\circ}$ and $\beta \approx 0.96$. Although these values 
are somewhat uncertain in the case of
Mrk~501, since this superluminal velocity was determined for a weak
feature whose position is only poorly defined, the joint analysis of 
transverse polarization profiles and apparent superluminal speeds provides
a new tool for disentangling $\delta$ and $\beta$. The jet
of Mrk~501 must be fairly close to the line of sight, yet carrying out this
analysis for the typically subluminal speeds observed in the Mrk~501 VLBI
jet yields viewing angles of about $60^{\circ}$; this essentially demonstrates 
that these subluminal motions must represent pattern speeds in a much more
highly relativistic flow, rather than highly relativistic motion viewed at a very
small angle.

Calculation of $I$ and $Q$ profiles for the helical-field model is greatly 
simplified if the spectral index $\alpha$ is assumed to be unity. Whilst this is 
almost certainly incorrect (observed values are usually $\simeq 0.5$) the
assumption that $\alpha = 1$ has been shown to have very little impact
on the profiles, and hence on the values of the best-fit model parameters.

The results demonstrate that this method provides a new approach to studying the
magnetic fields in parsec-scale jets. 
We are in the process of identifying other AGN jets that are well resolved and
straight, display clear transverse polarization structure,  
and possess reliable component speed measurements, which we hope will prove 
fruitful subjects of analyses similar to those carried out here. 

\vspace{-15mm}

\section{Acknowledgments}
Funding for this research was provided by the Irish Research Council for Science 
Engineering and Technology (IRCSET). We would also like to thank Andreas Papageorgiou 
for his previous work in this area and for useful discussions, and the anonymous 
referees for comments which we believe have helped make the paper more complete 
and clear.

\end{document}